\documentclass[twocolumn]{aastex631}

\shorttitle{A candidate location for Planet~9}
\shortauthors{Socas-Navarro}
\graphicspath{{./}{figures/}}

\begin{document}

\newcommand{\sone}{s$^{-1}$}
\title{A candidate location for Planet~9 from an interstellar meteoroid: The messenger hypothesis}

\author[0000-0001-9896-4622]{Hector Socas-Navarro}
\affiliation{Instituto de Astrof\'\i sica de Canarias,
  V\'\i a L\'actea S/N, La Laguna 38205, Tenerife, Spain}
\affiliation{Departamento de Astrof\'\i sica, Universidad de La Laguna, 38205, Tenerife, Spain }



\begin{abstract}

The existence of a hypothetical Planet~9 lurking in the outer solar
system has been invoked as a plausible explanation for the anomalous
clustering in the orbits of trans-Neptunian objects. Here we propose
that some meteoroids arriving at Earth could serve as messengers with
the potential of revealing the presence of a hitherto undiscovered
massive object. The peculiar meteor CNEOS 2014-01-08, recently put
forward as the first interstellar meteor, might be one such
messenger. The meteor radiant is in the maximum probability region
calculated for the Planet 9 location in previous works. The odds of this
coincidence being due to chance are~$\sim$1\%. Furthermore, some
statistical anomalies about CNEOS 2014-01-08 are resolved under the
hypothesis that it was flung at Earth by a gravitational
encounter. Integrating its trajectory backwards in time would then
lead to the region of the sky where Planet~9 is more likely to
reside.  Based on the available data, we propose the region at
coordinates
R.A.~53.0$\pm$4.3\textdegree , declination~9.2$\pm$1.3\textdegree \ 
as a plausible candidate location for Planet~9.
\end{abstract}




\section{Introduction}
 
The peculiar clustering of orbital parameters observed in very
eccentric extreme trans-Neptunian objects (ETNOs) has prompted the
suggestion that there might be a large planetary mass object still
undiscovered in the outer solar system
(\citealt{TS14,dFMdFM14,BB16a,BB16b,ML17}). These ETNOs are too far
away from Neptune to feel its gravitational influence. However, most
of them reach their perihelion in the ecliptic and have their
elliptical orbits aligned, with an ascending node ($\Omega$) and
orbital plane inclination ($i$) that are tightly clustered. This
coincidence could be explained by the presence of a large, very
distant planet with a mass between 4 and 8.4 times that of the Earth
and a semimajor axis between 300 and 520~AU (\citealt{BB21}, hereafter
BB21, see also \citealt{dFMdFM16}). It has even been suggested that
the unseen perturber might not be a planet at all but a primordial
black hole (\citealt{SJ20}), a hypothetical class of black holes
formed in the early Universe.

It should be noted that the evidence for ETNO clustering is not
undisputed. Some controversy exists on whether it is a real effect or
a statistical artifact caused by observational biases (see
\citealt{SKM+17,BBS+20,NGD21,BB21}). An independent clue supporting the
existence of a putative Planet~9 comes from the ETNO pair
2004~VN$_{112}$ and 2013~RF$_{98}$.  \cite{dLdlFMdlFM17} present
spectroscopic and dynamic evidence indicating that these two objects
were probably a binary asteroid that became detached near their
aphelion by a gravitational encounter with an unknown trans-Neptunian
planetary body.

The reader can find a review on our current knowledge of
the outer solar system in \cite{GK21}. If Planet~9 indeed exists, it
would be extremely difficult to detect and identify. First, it would
be a very faint object (probably around magnitude 20$\pm$2 in the
R~band). Furthermore, its orbital velocity would be extremely slow,
completing an orbit every 7,000 to 15,000~years (see BB21).

Multi-messenger astronomy has been gaining importance very rapidly
during the last decade and is often recognized as a high priority for
funding agencies (see, e.g., Astro2020,
\citealt{DecadalSurvey21}). The term messenger here means particles or
waves, other than photons, sent in our direction by celestial
bodies. Normally, this refers to neutrinos, gravitational waves and
cosmic rays. We note that small asteroids or meteoroids, accelerated
by massive bodies and eventually hitting Earth as meteors, behave as
particles that bring us information from distant astrophysical
objects. Thus, they may be considered as cosmic messengers, as
well. In this paper we argue that the first detection of such
meteoroid messenger could be CNEOS 2014-01-08 (hereafter CNEOS14),
perhaps unveiling to us the location of Planet~9.

\section{The meteoroid}
\label{sec:OSS}

In a search to discover new interstellar objects, \citet{SL22}
(hereafter SL22) parsed the \citeauthor{CNEOSdatabase} for meteors
with a hyperbolic trajectory.  Amongst all recorded events, SL22 found
one meteor with a clearly hyperbolic trajectory, CNEOS14, which
impacted at 17:05:34 (UT) near Papua New Guinea (impact coordinates
1.3\textdegree S, 147.6\textdegree E).  Its heliocentric velocity at
that moment was $\simeq$60~km~\sone , according to SL22, or
61$\pm$5.8~km~\sone , according to \citet{PATRR22}.

\begin{figure*}
    \centering
    \includegraphics[width=0.8\textwidth]{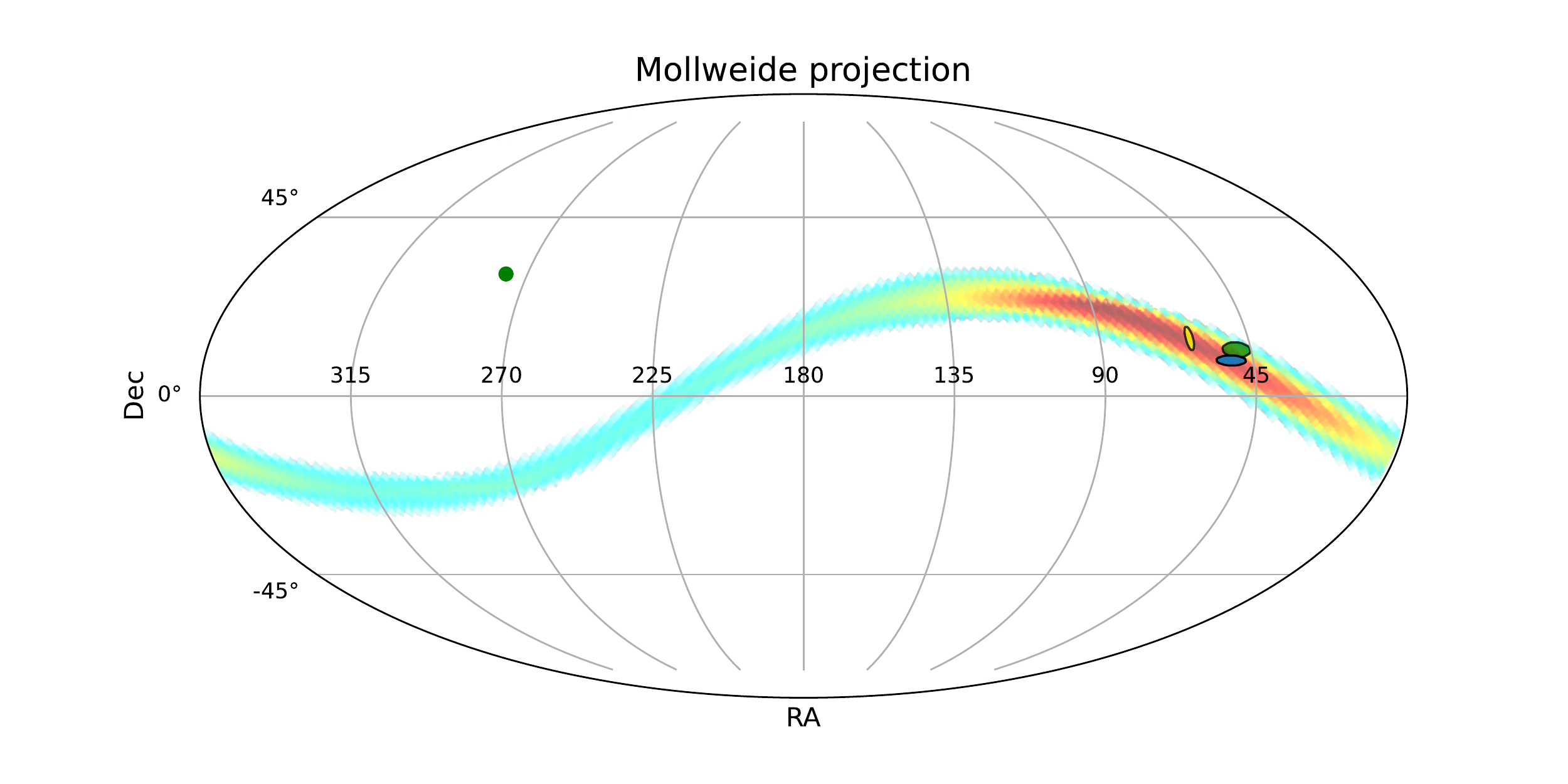}
    \caption{Map of the entire sky in a Mollweide projection. The
      colored band marks the range of possible Planet~9 orbits, as
      obtained by BB21 (greater probability in red colors).
      Horizontal blue ellipse: CNEOS14 coordinates when it was at a
      distance between 300 and 600~AU from the Sun (the predicted
      range of distances for the Planet~9 orbit). Horizontal green
      ellipse: Asymptotic radiant (i.e., at infinity) obtained by
      SL22. Vertical yellow ellipse: Geocentric radiant (i.e.,
      immediately before impact) obtained by \cite{PATRR22}. The green
      dot marks the solar apex, i.e. the direction of motion of the
      solar system.
    \label{fig:figmap}
    }
\end{figure*}

By means of numerical simulations (detailed below), we traced back the
trajectory of CNEOS14 to the outer solar system and
determined the meteoroid position in the sky (distant radiant) as it
traversed the Planet~9 possible orbits, at distances between 300 and
600~AU from the Sun. The distant radiant obtained was
R.A.~51.6$\pm$4.3\textdegree, dec~8.6$\pm$1.2\textdegree . Strikingly,
this location is within the band of possible Planet~9 locations
published in BB21, as shown in Figure~\ref{fig:figmap}. In fact it lies
well within the red area of higher probability to find the
planet. This is a very remarkable coincidence since, as we discuss
below, it is rather unlikely to occur by chance.

Other radiant coordinates given in previous works have slightly
different meanings, which might account for the small discrepancies
amongst them (in addition, of course, to possible differences in the
simulation details). \cite{PATRR22} obtained
R.A.~62.8$\pm$1.1\textdegree, dec~14.0$\pm$2.8\textdegree \, for the
geocentric radiant. This refers to the incoming directorion of the
meteoroid before the impact if its trajectory had not been curved by
Earth's gravity. SL22 obtained R.A.~49.4$\pm$4.1\textdegree,
dec~11.2$\pm$1.8\textdegree \, for the asymptotic radiant, which is
the direction of the meteoroid at infinity. 

We explored the trajectory of CNEOS14 by integrating its motion
backwards from the impact location throughout the solar system, out to
1000~AU. A newtonian N-body simulation code was employed, accounting
for the gravitation of the Sun, the planets and the Moon. This code is
publicly available and has been described elsewhere (\citealt{SN19}). 

The simulation is initialized with the meteoroid at the impact site
and observed velocity. The location and initial velocities for the
planets and the Moon are obtained from the JPL Horizons system
(\citealt{Horizons_G15}). A total of 1,000 clones are generated with
random perturbations applied to the initial velocity in order to
account for the observational uncertainties. In the first 500 clones,
the initial velocity perturbation follows a Gaussian distribution with central value 0 and standard deviation equal to 10\% of the observed value.

The remaining 500 clones have their initial velocity modulus multiplied by a
random factor uniformly distributed between 1.0 and 1.1 to account for
the possibility that the meteoroid might have experienced some
deceleration (up to 10\%) upon contact with the atmosphere.

The simulation is divided into two stages. In the first stage, the
meteoroid is very close to Earth and the Moon, and we need a higher
temporal resolution to obtain an accurate trajectory
reconstruction. For this phase we use a time resolution of 1~s and
compute one million time-steps. The simulation time spanned in this phase
is 11.57 days and ends with the meteoroid in interplanetary space at
0.3~AU from Earth. The second phase starts at the end of
phase 1, has a coarser time resolution and a much longer span of
simulation time to cover the meteoroid trip into the outer solar
system. For this phase we used a time resolution of 1~hour and again
one million time-steps, spanning a total of 110 years of simulation time.

\begin{figure}
    \includegraphics[width=0.5\textwidth]{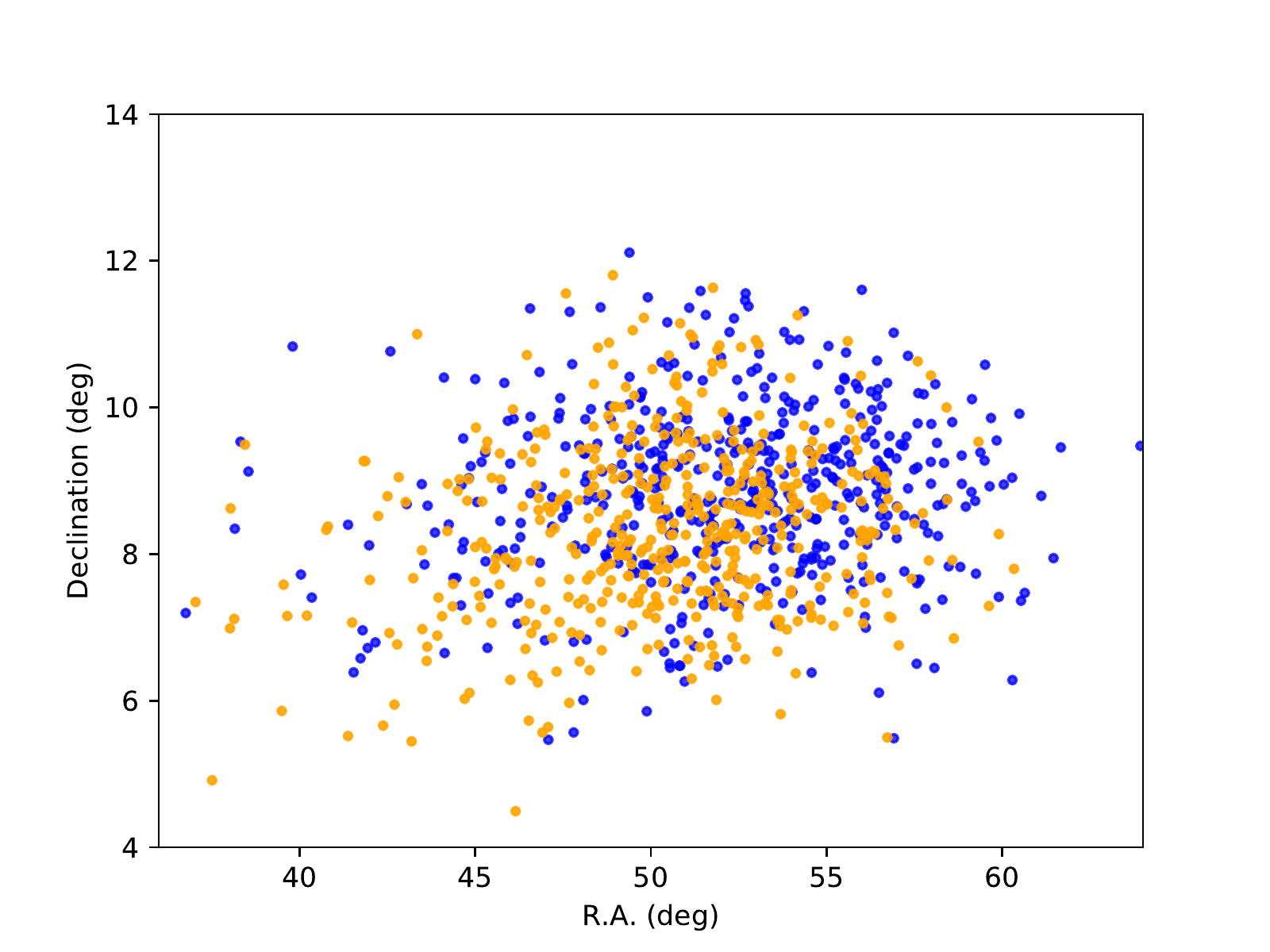}
    \caption{Distribution of simulation clones at 300~AU. Orange dots:
      First 500 clones, no atmospheric slowdown effect
      considered. Blue dots: Remaining 500 clones with initial
      velocities augmented by a random factor between 1 and 1.1
      (uniform distribution).
    \label{fig:coords}
    }
\end{figure}

According to the simulations, the meteoroid crossed the 300 and 600~AU
heliocentric distances 32 and 65~years before the impact.  Its
celestial coordinates (radiant) did not vary significantly between 300
and 600~AU but they do vary among clones. They are plotted in
Figure~\ref{fig:coords}, with orange dots representing the first 500
clones and blue dots representing the clones with the atmospheric
deceleration of up to 10\%. The median coordinates are
R.A.~50.7$\pm$4.1\textdegree , declination 8.3$\pm$1.2\textdegree
\ for the first 500 clones and R.A.~51.6$\pm$4.3\textdegree ,
declination ~8.6$\pm$1.2\textdegree \ when the entire sample is
considered.  Note that the errors on these coordinates are not
uncorrelated. On average, clones with higher values of R.A. have
slightly higher declinations, as well.
   
\section{Statistical anomalies}
\label{sec:anomalies}

There are some intriguing statistical anomalies about 
CNEOS14 if we consider it as an interstellar object with an
unperturbed inbound trajectory from the interstellar medium.

\begin{enumerate}
\item Its approach trajectory was very close to the plane of the
  ecliptic, with only a 7\textdegree \, inclination. An interstellar
  object would be expected to come in from any random
  direction. Assuming an isotropic distribution of inbound velocities,
  the inclination angles ($i$) with respect to the plane of the
  ecliptic would follow a uniform distribution in $\sin(i)$ and the
  cumulative probability for a single object to have an inclination
  lower than $i_0$ would be given by $P(i<i_0) = \sin(i_0)$. In our
  case, $P(i < 7$\textdegree$)$ is only 12\%. We can compare this
  value to the other known interstellar interlopers. 1I/'Oumuamua had
  an inclination of 122\textdegree \, (i.e., 58\textdegree \ from the
  ecliptic plane but moving in a retrograde motion), resulting in a
  cumulative probability $P$=85\%. 2I/Borisov had a 44\textdegree \,
  inclination, yielding a probability $P$=69\%. Of course, the current
  sample of interstellar interlopers is too limited for a meaningful
  comparison. They are mentioned here simply as illustrative examples
  exhibiting the expected statistical behavior.

\item Its velocity with respect to the {\it local standard of rest}
  (LSR, the mean motion of matter around the galactic center) was
  oddly large at $v_{LSR}$=58~km~\sone \ (SL22).  The 1-$\sigma$
  velocity dispersion in the solar neighborhood is $\sim$38~km~\sone \ 
  (\citealt{DB98}), which means that only 6\% of stars move at
  58~km~\sone \ or faster (as a reference, the Sun moves with respect
  to the LSR at 18~km~\sone, \citealt{SBD10}; \citealt{ZS19}).  The
  LSR speed of CNEOS14 is then peculiarly high.  For comparison with
  the other two known interstellar interlopers, the speeds of
  1I/'Oumuamua and 2I/Borisov with respect to the LSR were
  11~and~35~km~\sone , respectively (\citealt{Mamajek17};
  \citealt{BJCF+20}).  If we consider that CNEOS14 was ejected from a
  ``normal'' star moving slower than 38~km~\sone, it must have
  received a kick strong enough to eject it with a relative velocity
  at infinity v$_\infty$ of at least 20~km~\sone \ with respect to its
  parent star. To put this into context, the Voyager spacecraft, after
  two gravity assist maneuvers with Jupiter and Saturn, are leaving
  the solar system at v$_\infty$=17 km~\sone. Not only the kick would
  need to be extremely strong but also aligned with the direction of
  the star motion in order to eject CNEOS14 into the interstellar
  medium at a speed of 58~km~\sone \, with respect to the LSR. Such
  alignment would be extremely unlikely. SL22 suggest that perhaps the
  parent star is from the Milky Way thick disk, which have a larger
  velocity dispersion of $\sim$50~km~\sone. This is possible but again
  unlikely since the thick disk population constitutes only about
  $\sim$10\% \ of the stars. In summary, CNEOS14 had a peculiarly high
  velocity with respect to the LSR, compatible with only 6\% \ to 10\%
  \ of stars.

\item SL22 claim that, in order to detect one interstellar object in a
  period of $\sim$10~years, the number density of such objects in our
  environment should be $n=10^{6^{+0.75}_{-1.5}}$~AU$^{-3}$ (at 95\%
  confidence, assuming Poisson statistics) and such amount of objects
  would be at tension with the mass of available planetesimal material
  inside the 60~km~\sone \ orbit expected in a solar nebula. This
  means that, with our current knowledge, it is unlikely that we would
  have an interstellar object like CNEOS14 in the database. The
  identification of interstellar meteors is a very challenging task
  with a long history of investigations (see e.g., \citealt{O50, W14,
    HSW+20}). Given these difficulties it is likely that some events
  might have gone unnoticed, implying that the actual number (and the
  tension) would be even greater.
\end{enumerate}

When considered together, the probability that these three anomalies
are due to chance is less than 1\%. In the remainder of this paper, we
discuss a hypothetical scenario that resolves these anomalies by
postulating that the trajectory of CNEOS14 was deflected by a
gravitational encounter with a massive body in the outer solar system.

\section{Direct impactor or messenger}

There are two possibilities regarding the origin of CNEOS14. The first
one is that it was an interstellar asteroid that entered the solar
system in a direct collision course with Earth and hit us without the
mediation of any other intervention (the direct impactor
hypothesis). The other possibility, that we put forward in this paper,
is that it experienced a gravitational encounter with a massive body
and was slingshotted in our direction (the messenger 
hypothesis).

The direct impactor hypothesis does not explain the remarkable
coincidence between the Planet~9 high-probability region and the
meteoroid radiant, highlighted in Figure~\ref{fig:figmap}. Such
coincidence is rather unlikely to occur by chance. If the meteoroid
entered the solar system from any random incoming direction, it would
have equal probability of originating anywhere in the figure (the
Mollweide projection preserves areas). The probability of coincidence
by chance with the high-probability (red) contour is~1\% (the ratio of
solid angles subtended by the contour and the entire sky). If we
consider instead the entire Planet~9 band around the sky calculated by
BB21, then the probability of a chance coincidence would increase to
9\%.

The probabilities quoted above are upper bounds because they assume,
for simplicity, an isotropic distribution of velocities for
interstellar objects in the heliocentric frame. The solar system is
moving at 18~km~\sone \ with respect to the LSR (\citealt{SBD10};
\citealt{ZS19}) in the direction of the solar apex, marked by a green
dot in the figure. This means that the probability of interstellar
asteroids entering from that side of the sky will be relatively higher
and, conversely, lower from the other side. The solar apex and the
CNEOS14 radiant are separated in R.A. by 133\textdegree \ and therefore
the number of objects entering the solar system from that direction
should be lower, thus lowering the probability of chance coincidence
quoted in the previous paragraph.

The messenger hypothesis, on the other hand, explains the coincidence
of the CNEOS14 radiant with the Planet~9 high probability contour and
also resolves the CNEOS14 anomalies discussed in
Section~\ref{sec:anomalies} above. There would be nothing particularly
peculiar about CNEOS14. A significant part of its large impact
velocity of 60~km~\sone \ in the heliocentric frame is provided by the
solar gravitational acceleration (up to 42~km~\sone). The remaining
kinetic energy, corresponding to another 42~km~\sone , would be the
combination of its inbound velocity from the interstellar medium in
the heliocentric frame plus the delta-v supplied by the Planet~9
slingshot. In the limiting case of zero delta-v (i.e., the slingshot
would have only caused a deflection but no velocity increase), the
required 42~km~\sone \ is very close to the typical rms velocity of
objects with respect to the LSR (38~km~\sone ). For other delta-v
values, the meteoroid velocity in the LSR would be even lower.

Thus far we have discussed CNEOS14 in the context of previous Planet~9
clues. One might also consider what this meteor tells us independently
of all previous work. CNEOS14 was found in a search for a meteor with
high heliocentric speed, returning a match with $\sim$60~km~\sone . At
such a high speed, CNEOS14 was not bound to the solar gravity and
therefore it must be either an interstellar direct impactor or
a messenger accelerated into our direction by a gravitational
encounter.

In the first case (direct impactor), the meteor would be expected to
arrive from any direction in space with a random inclination (uniform
in $\sin(i)$) with respect to the plane of the ecliptic. However, the
trajectory followed by CNEOS14 was very close to the ecliptic, with
only a 7\textdegree \ inclination. There is only a 12\% probability for
a chance alignment, which disfavors this first scenario.

Furthermore, if CNEOS14 were an undeflected meteoroid, its speed with
respect to the LSR would be of 58~km~\sone \ which, as
discussed in Section~\ref{sec:anomalies}, is peculiarly high. Only
about 6\% of stars move that fast.

Combining these two probabilities as independent (uncorrelated)
properties, the odds that CNEOS14 had a gravitational encounter during
its passage through the solar system before hitting Earth may be
estimated conservatively to be higher than 99\%. However, it did not
come close to any of the known planets. Table~\ref{table:planets}
lists the approximate dates at which the projected trajectory of
CNEOS14 crossed each one of the outer planet orbits and the
coordinates of the planets at that time.

\begin{table}
\caption{Coordinates of outer planets on the date when CNEOS14 crossed their orbits. None of them was close in the sky to the meteoroid (approx. R.A.: 51.6\textdegree, dec: 8.6\textdegree)}             
\label{table:planets}      
\centering                          
\begin{tabular}{c c c c}        
\hline\hline                 
Planet  & Orbit crossing & R.A. & dec \\    
\hline                        
Mars  &  2013-12-21  &  187.24\textdegree  &  0.88\textdegree  \\
Jupiter  &  2013-08-12  &  101.33\textdegree  &  22.88\textdegree  \\
Saturn  &  2013-03-01  &  219.80\textdegree  &  -11.19\textdegree  \\
Uranus  &  2012-02-19  &  2.71\textdegree  &  0.41\textdegree  \\
Neptune  &  2010-12-13  &  328.70\textdegree  &  -12.79\textdegree  \\
\hline                                   
\end{tabular}
\end{table}

Therefore, even if one disregards all previous evidence for Planet~9
based on the ETNO clustering, this meteor is most likely a messenger
slingshotted in our direction by some unknown massive body in the
outer solar system or beyond. This is additional evidence in favor of
the existence of Planet~9 other than the one provided by the peculiar
orbital parameter space of the ETNOs.

The current position of Planet~9 has changed in the 32 to 65~years
elapsed since the hypothetical encounter with the meteoroid because of
the planet's proper motion. Given the orbital parameters estimated for
Planet~9 (e.g. BB21), the proper motion must be between 1.3\textdegree
\ and 1.7\textdegree . Taking this into account, we end up with
coordinates
R.A.~53.0$\pm$4.3\textdegree , declination~9.2$\pm$1.3\textdegree \      
as the predicted current position for Planet~9.

\section{Conclusions}

We propose a plausible current candidate location for Planet~9 at
coordinates R.A.~53.0$\pm$4.3\textdegree ,
declination~9.2$\pm$1.3\textdegree \ , in the trijunction of Aries,
Taurus and Cetus.  Moreover, we argue that CNEOS14 provides additional
supporting evidence in favor of Planet~9 that is completely
independent of the distant ETNOs orbital clustering. It should be
emphasized that the proposed location is only a candidate, there is no
certainty about this prediction, which could be just a curious
coincidence. However, it is a plausible, easily testable and well
motivated proposal. The odds of a chance coincidence between
Planet~9's orbit and the CNEOS14 origin is lower than 9\% or 1\%
(depending on whether one considers the full orbital band or the
higher probability region). Figure~\ref{fig:figmap} shows these
locations, as well as the direction of the solar system motion (the
solar apex), which is on the opposite side of the sky. The statistical
significance of this coincidence is not very high but it increases
considerably when combined with the statistical anomalies of CNEOS14
for an overall $\sim$0.1\% probability of chance.

There are two possibilities: Either CNEOS14 was a direct impactor or
it was redirected into our direction by a gravitational encounter
(messenger hypothesis).

The messenger scenario explains the coincidence of the CNEOS14 radiant
with the Planet~9 orbital band and resolves the CNEOS14 anomalies in
Section~\ref{sec:anomalies}. The meteoroid would not need to be moving
initially at 58~km~\sone \ with respect to the LSR. It could have been
moving at typical speeds of stars in the LSR.

We consider the messenger hypothesis as the most parsimonious with all
existing evidence. There are, however, a number of caveats that should
be kept in mind:

\begin{enumerate}
  \item Planet~9 might not exist. The current evidence is indirect and
    controversial. The ETNO clustering has been attributed by some
    authors to observational bias (\citealt{SKM+17,BBS+20,NGD21}). For
    counterarguments, see \cite{BB21}. Nevertheless, there is
    additional support for the Planet~9 hypothesis beyond the ETNO
    clustering. \cite{dLdlFMdlFM17} argue that the pair of ETNOs
    2004~VN$_{112}$ and 2013~RF$_{98}$ originated as a binary asteroid
    and were detached by a trans-Neptunian planetary
    encounter. Furthermore, in this paper we find a remarkable
    coincidence between the CNEOS14 radiant and the region where
    Planet~9 is expected to reside that would be rather unlikely to
    occur by chance.
  \item Most of the argumentation presented in this paper is based on
    the determination that the CNEOS14 trajectory is hyperbolic, given
    its high heliocentric speed at 1~AU. If such determination were
    deemed incorrect (as suggested by \citealt{V22}), then the
    meteoroid could be a solar system object and, in that case, the
    statistical significance of its anomalies would decrease
    enormously. However, the observational evidence for the hyperbolic
    trajectory is very robust. As pointed out by SL22, in order to
    make it a bound orbit the measured velocity would need to be in
    error by about 20~km~\sone , or nearly 50\%. Independent studies
    of the accuracy of events in the CNEOS database have not observed
    such large errors (\citealt{DBS+19}; \citealt{BSG+17};
    \citealt{PATRR22}).
  \item Even if the messenger hypothesis is correct, the massive
    perturber could be something other than Planet~9. One possibility
    is that the meteoroid originated in the Oort cloud during a close
    encounter with a passing star (perhaps even in that star's Oort
    cloud). We explored this hypothesis by comparing the timing of the
    meteoroid origin to the list of recent stellar
    encounters. Extending the simulation out to distances beyond
    $\sim$1~lyr, we find that such distance would have been reached
    7~kyr ago (24~kyr for 1~pc). However, no such encounters closer
    than 1~pc have occurred more recently than 80~kyr ago
    (\citealt{BJ22}; \citealt{M15}). Another possibility could be an encounter with a
    passing brown dwarf or a rogue planet (meaning a free-floating
    planet in the interstellar medium) thousands of years ago, again
    destabilizing objects in the Oort cloud. This scenario is
    certainly a possibility but it would not explain the coincidence
    of the meteoroid with the predicted location for Planet~9 or the
    low ecliptic inclination.
  \item \cite{HGB+22} propose an alternative explanation in which the ETNO
    population was created by a distant planet that is not there any
    more. In their scenario, Planet~9 would have been a temporary
    inhabitant of the solar system (a rogue planet in their
    terminology, not to be confused with the meaning of rogue in the
    previous paragraph) and our efforts to understand the
    trans-Neptunian peculiarities would be planetary archaeology. This
    scenario is consistent with all the available evidence except for
    the clues presented in this paper, which require the presence of a
    solar system Planet~9 between 32 and 65 years ago.
  \item An open question in the messenger hypothesis is why would the
    first interstellar meteor be a messenger and not a direct
    impactor. The answer might be related to the search bias. This
    object was identified by searching for the highest heliocentric
    speed in a database of meteor events. Perhaps the fastest objects
    are those resulting from a gravitational slingshot and that
    introduces a bias in the search by making them more easily
    recognizable amongst all background events. This issue may be explored
    with detailed simulations of the solar system motion through the
    interstellar medium. A natural follow up question could be why
    this messenger comes from Planet~9 and not Jupiter or the other
    giant planets, which are considerably more massive. However, the
    sphere of gravitational influence, given by the Hill radius,
    depends much more strongly on the distance to the Sun than the
    planetary mass. While Jupiter has a Hill radius of 0.35~AU,
    Planet~9 has it between 5 and 12~AU. Thus, the Planet~9 volume of
    gravitational dominance is between 2,400 and 38,600~times larger
    than Jupiter's.
  \item Perhaps the most important caveat is that the Planet~9
    location proposed here is sensitive to the accuracy of the CNEOS14
    observed velocity. Previous studies comparing events in the CNEOS
    database with ground-based observations found that, while a good
    agreement is obtained in most cases, significant discrepancies are
    found in a few events (\citealt{DBS+19}; \citealt{BSG+17};
    \citealt{PATRR22}). Unfortunately, no other observations exist of
    CNEOS14, so we need to rely on this single measurement.
\end{enumerate}

With all of these considerations in mind, the evidence in favor of the
Planet~9 messenger hypothesis for the origin of CNEOS14 seems
sufficiently compelling to warrant a search effort. The slow proper
motion makes the identification of Planet~9 extremely challenging in
observations. It takes a very long time to detect a significant motion
with respect to the background stars. However, parallax observations
might prove helpful. At the predicted distances, a parallax between
12\arcmin \ and 23\arcmin \ is expected. Unfortunately, the BB21
calculations suggest that Planet~9 is just beyond the magnitude limit
for GAIA. However, given the exploratory nature of their calculation
and the marginal difference with the limit, it is probably worth some
search efforts.

\begin{acknowledgements}
This research has made use of NASA's Astrophysics Data System
Bibliographic Services, as well as the Jet Propulsion Laboratory's
CNEOS and Horizons databases. Python Matplotlib (\citealt{H07}),
Numpy (\citealt{numpy11}) and iPython (\citealt{ipython07}) modules have been
employed to generate the figures and analyze the calculations in this
paper.  Thanks are also due to Eloy Peña-Asensio for many fruitful
discussions.
\end{acknowledgements}

\bibliographystyle{aa}
\bibliography{aanda.bib,paper.bib}

%
%

\end{document}